%% file: final.tex
\documentstyle[epic,eepic,fullpage]{article}

\newcounter{level1ctr}
\newenvironment{level1}{\begin{list}{\arabic{level1ctr}.}{
        \usecounter{level1ctr}
        \settowidth{\labelwidth}{8.}
        \addtolength{\labelsep}{-2pt}
        \setlength{\leftmargin}{\labelwidth}
        \addtolength{\leftmargin}{\labelsep}}}{\end{list}}

\begin{document}
\date{}
\title{\Large\bf Secure Execution of Java Applets using a Remote Playground\thanks{Preprint of a paper to appear in IEEE Transactions on Software
Engineering.}}
\author{\begin{tabular}[t]{c@{\extracolsep{8em}}c}
Dahlia Malkhi  & Michael K.\ Reiter \\
\\
AT\&T Labs Research & Bell Laboratories \\
Florham Park, NJ, USA & Murray Hill, NJ, USA \\
{\tt dalia@research.att.com} & {\tt reiter@research.bell-labs.com}
\end{tabular}}
\maketitle

\subsection*{\centering Abstract}
{\em Mobile code presents a number of threats to machines that execute
it.  We introduce an approach for protecting machines and the
resources they hold from mobile code, and describe a system based on
our approach for protecting host machines from Java 1.1 applets.  In
our approach, each Java applet downloaded to the protected domain is
rerouted to a dedicated machine (or set of machines), the {\em
playground}, at which it is executed.  Prior to execution the applet
is transformed to use the downloading user's web browser as a graphics
terminal for its input and output, and so the user has the illusion
that the applet is running on her own machine.  In reality, however,
mobile code runs only in the sanitized environment of the playground,
where user files cannot be mounted and from which only limited network
connections are accepted by machines in the protected domain.  Our
playground thus provides a second level of defense against mobile code
that circumvents language-based defenses. The paper presents the
design and implementation of a playground for Java 1.1 applets, and
discusses extensions of it for other forms of mobile code including
Java 1.2.
}

\vspace{0.5in}
\noindent
{\bf Keywords: }
Java, mobile-code, security, remote method invokation.

\section{Introduction}
\label{intro}

Advances in mobile code, particularly Java, have considerably
increased the exposure of networked computers to attackers.  Due to
the ``push'' technologies that often deliver such code, an attacker
can download and execute programs on a victim's machine without the
victim's knowledge or consent.  The attacker's code could conceivably
delete, modify, or steal data on the victim's machine, or otherwise
abuse other resources available from that machine.  Moreover, mobile
code ``sandboxes'' intended to constrain mobile code have in many
cases proven unsatisfactory, in that implementation errors enable
mobile code to circumvent the sandbox's security
mechanisms~\cite{DFW96,MF97}.

One of the oldest ideas in security, computer or otherwise, is to
physically separate the attacker from the resources of value.  In this
paper we present a novel approach for physically separating mobile
code from those resources.  The basic idea is to execute the mobile
code somewhere other than the user's machine, where the resources of
value to the user are not available, and to force the mobile code to
interact with the user only from this sanitized environment.
The challenge is to achieve this physical separation without
eliminating the benefits derived from code mobility, in particular
reducing load on the code's server and increasing performance by
co-locating the code and the user.

In order to achieve this protection at an organizational level, we
propose the designation of a distinguished machine (or set of
machines), a {\em playground}, on which all mobile code served to a
protected domain is executed.  That is, any mobile code pushed to a
machine in this protected domain is automatically rerouted to and
executed on the domain's playground.  To enable the user to interact
with the mobile code during its execution, the user's computer acts as
a graphics terminal to which the mobile code displays its output and
from which it receives its input.  However, at no point is any mobile
code executed on the user's machine.  Provided that valuable resources
are not available to the playground, the mobile code can entirely
corrupt the playground with no risk to the domain's resources.  Our
playground thus provides a second level of defense against mobile code
that circumvents language-based defenses.  Moreover, because the
playground can be placed in close network proximity to the machines in
the domain it serves, performance degradation experienced by users is
minimal.  There can even be many playgrounds serving a domain to
balance load among them.

In this paper we report on the design and implementation of a
playground for Java 1.1 applets.  As described above, our system
reroutes all Java applets retrieved via the web to the domain's
playground, where the applets are executed using the user's browser
essentially as an I/O terminal.  This approach enjoys two advantages:
(1) The playground is centrally controlled by a security administrator
and can be reinforced for security, e.g., by a secure operating system
or with various add-on security tools; consequently it is
not sensitive to a blunder that may be caused by a user setting a
security policy in a browser or to weaknesses exposed in less secure
configurations in the network. (2) At the same time, the domain's
resources can be protected even if the playground is completely
corrupted, by disallowing the playground to mount protected file
systems or open arbitrary network connections to domain machines---in
the limit, locating the playground just ``outside'' the domain's
firewall. Our system is largely transparent to users and applet
developers, and in some configurations requires no changes to web
browsers in use today.  While there are applets that are not suited to
execution on our present playground prototype, e.g., due to
performance requirements or code structure, in our experience these
are a small fraction of Java applets.

As described above, the playground need not be trusted for our system
to work securely.  Indeed, the only trusted code that is common to all
configurations of our system is the browser itself and a small
``graphics server'', itself a Java applet, that runs in the browser.
The graphics server implements interfaces that the untrusted applet,
running on the playground, calls to interact with the user.  The
graphics server is a simply structured piece of code, and thus should
be amenable to analysis.  In one configuration of our system, trust is
limited to {\em only} the graphics server and the browser, but doing
so requires a minor change to browsers available today.  If we are
constrained to using today's browsers off-the-shelf, then a web proxy
component of our system, described in Section~\ref{arch}, must also be
trusted.

The rest of this paper is structured as follows.  In
Section~\ref{related} we review related work.  We describe the
architecture of our system in Section~\ref{arch}, and discuss its
security in Section~\ref{security}. We provide implementation details
and examples in Section~\ref{implementation}.  Transparency of applet
execution is discussed in Section~\ref{transparency}. We discuss the
extension of our approach to Java 1.2 in Section~\ref{1.2} and
conclude in Section~\ref{conclusion}.

\section{Related work}
\label{related}

There are three general approaches that have been previously proposed
for securing hosts from mobile code.  The first to be deployed on a
large scale for Java is the ``sandbox'' model.  In this model, Java
applets are executed in a restricted execution environment (the
sandbox) within the user's browser; this sandbox attempts to prevent
the applet from performing illegal actions.  This approach has met
with mixed success, in that even small implementation errors can
enable applets to entirely bypass the security restrictions enforced
by the sandbox~\cite{DFW96}. Several systems were built that reinforce
the JVM and enhance the sandbox security model in various ways.  These
include add-on commercial solutions that enhance the security of the
JVM through additional mechanisms installed in the
environment. Finjan's SurfinShield (see {\tt www.finjan.com}) enhances
standard browsers with utilities to monitor and manage mobile
code. eSafe's Protect (see {\tt www.esafe.com}) provides an additional
sandbox to run applets in, which places resource limitations on
applets. Security7's SafeAgent (see {\tt www.security7.com}) isolates
active code on a workstation from any other executables and resources.
Some of these products are coupled with applet filtering mechanism's
at the firewall (see below).  Other efforts explore extending the JVM
into a more fully capable language-oriented (Java) operating
system. Alta and GVM~\cite{BTS+98} and the J-Kernel/JRes
system~\cite{CvE98,HCG+98} are examples of Java operating systems
supporting multiple Java applications and providing enhanced resource
protection and management among them.  Some of the issues concerning
multi-processing in Java are explored
in~\cite{BG98}. Joust~\cite{HPB+97} is a Java operating system
exploiting the {\em paths\/} abstraction in Scout~\cite{MP96} to
provide higher degree of control and security of Java interaction with
its environments. These systems can better separate Java applications
co-running on a JVM (e.g., on low-end devices) as well as provide
better control and monitoring of Java applet behavior. Additionally, a
vast amount of previous research is dedicated to operating systems
security mechanisms in general. This research is beyond the scope of
this paper.

The second general approach is to execute only mobile code that is
{\em trusted} based on some criteria.  For example, Balfanz and Felten
proposed a {\em Java filter} that allows users to specify the servers
from which to accept Java applets~\cite{BF97}.  Here the criterion by
which an applet is trusted is the server that serves it.  A related
approach is to determine whether to trust mobile code based on its
author, which can be determined, e.g., if the code is digitally signed
by the author.  This is the approach adopted for securing Microsoft's
ActiveX content, and is also supported for applets in JDK 1.1.
Combinations of this approach and the sandbox model are implemented in
JDK 1.2~\cite{Gon97,GMPS97} and Netscape Communicator
(see~\cite{WBDF97}), which enforce access controls on an applet based
on the signatures it possesses (or other properties).  A third
variation on this theme is {\em proof-carrying code}~\cite{NL96},
where the mobile code is accompanied by a proof that it satisfies
certain safety properties.  This approach moves the intensive portions
of checking memory safety and type safety of mobile code expressed in
unsafe languages to the code producer, and it has been demonstrated in
several settings~\cite{N97}.  However, proof-carrying code is limited
in its ability to prove arbitrary security properties of code, and in
the case of Java, is less essential because the memory and type safety
properties of Java bytecodes can be efficiently checked by the code
consumer.  (On the other hand, flaws in these checks have been a
source of security vulnerabilities of Java~\cite{DFW96}.)

Our approach is compatible with both of the approaches described
above.  Our playground executes applets in sandboxes (hence the name
``playground''), which could be reinforced by various add-on
security tools. Likewise, it could easily be adapted to execute only
``trusted'' applets based on any of the criteria above.  Our approach
provides an orthogonal defense against hostile applets, and in
particular, in our system a hostile applet is still physically
separated from valuable resources after circumventing these other
defenses.

The third approach to securing hosts from mobile code is simply to not
run mobile code.  A course-grained approach for Java is to disable
Java in the browser.  Another approach is to filter out all applets at
a firewall~\cite{MRR97} (see also~\cite[Chapter 5]{MF97}), which has
the advantage of allowing applets served from behind the firewall to
be executed, but as pointed in~\cite{MRR97}, has several
limitations. Several companies provide products that attempt to
``blacklist'' potentially harmful applets at the firewall based on
various criteria or on content inspection, e.g., Finjan's SurfinGate
and SurfinCheck, Security7's SafeGate, and eSafe's Protect Gateway.

Independently of our work, a system with very similar goals and
architecture has recently been marketed by Digitivity, Inc., and
subsequently acquired by Citrix.\footnote{Digitivity's system is
subject to patent application in the U.S., U.K., and other parts of
the world.}  In Digitivity's terminology~\cite{Her97}, an AppRouter
redirects any applet loaded from the Internet onto a Cage machine,
which is designated to run applets while porting their GUI remotely to
the user's browser. The Cage differs from our playground in several
ways: First, the Cage JVM contains a graphics driver that reduces all
the Java GUI to a proprietary graphics protocol, and ports all GUI
from the Cage over a proprietary asynchronous communication protocol
onto the user's browser. Compared with our pure Java implementation,
the remote GUI for the Cage is less available for public scrutiny and
use, and is less portable, but has several notable advantages: First,
the proprietary communication protocol is tuned by Digitivity for
performance and security.  Second, their system is fully transparent,
whereas our design does not support a small (in our experience) class
of applets (see Section~\ref{transparency}).  Third, Digitivity's
system is developed to a marketable product level that our research
prototype has not reached. For example, the Cage itself contains
operating system reinforcement mechanisms to increase the security and
control of the Cage JVM. As another example, class file loading onto
the Cage bypasses the AppRouter to prevent a potential bottleneck
forming there.  Such mechanisms are compatible with our architecture
and can be added to it, but are outside the scope of our effort. On
the other hand, Digitivity's architecture relies on trusting AppRouter
for redirecting applets onto the Cage, whereas we have an alternative
that excludes this proxy component from our trusted base, as discussed
in Section~\ref{prevent-class-loading}.

The idea of allowing proxying graphics/windows interaction is itself
not new.  The most closely related tool to ours is Remote AWT
(RAWT)~\cite{RBW99}, recently developed by IBM; more information on
RAWT is available from {\tt www.alphaworks.ibm.com}.  Similar to our
design, RAWT is a Java implementation porting the AWT---i.e., the
standard API for implementing graphical user interfaces (GUI) in Java
1.1 programs---onto a remote Java graphics server.  RAWT is shipped as
a standalone tool supporting networked operation and resource sharing
for Java applications.  It can also be used for securely executing
Java applets, but will require additional modules interacting
specifically with browsers to automate applet redirection to some
sanitized environment, as in our proposal. RAWT's implementation
differs from ours in several respects, some of which are detailed in
Section~\ref{transparency} when we discuss transparency of applet
execution. More remotely related to our work are several prior tools
that enable X11 clients to interact with multiple and/or remote
displays~\cite{AbF91,MaN91}, support mobile X11 users~\cite{Ric95},
and even direct X11 connections across the Internet onto a Java
X11-display emulator~\cite{WRBHH96}. All of these tools rely on
platform specific X11 server modules.  We chose to implement our
remote graphics server in Java, and thus it is portable to any
Java-compliant environment, regardless of the host windows
environment.  Moreover, the connection between our graphics server and
the playground applet can be passed safely accross a packet-filtering
firewall, avoiding the security concerns associated with X11
connections~\cite[Ch.\ 3.3.3]{CB94}.

Finally, the idea of using a {\em sacrificial lamb machine\/} (also
termed a {\em bastion host\/}) to execute various services,
especially Internet services, has been used by numerous systems in the
past, specifically in firewalls (see~\cite[Ch. 5]{CZ95} for a
thorough treatment).

\section{Architecture}
\label{arch}

The core idea in this paper is to establish a dedicated machine (or
set of machines) called a {\em playground} at which mobile code is
transparently executed, using users' browsers as I/O terminals.  In
this section we give an overview of the playground and supporting
architecture that we implemented for Java, deferring many details to
Section~\ref{implementation}.

To understand how our system works, it is first necessary to
understand how browsers retrieve, load, and run Java applets.  When a
browser retrieves a web page written in Hypertext Markup Language
(HTML), it takes actions based on the HTML {\em tags} in that page.
One such tag is the {\tt <applet>} tag, which might appear as
follows:\\
\\ {\tt <applet code=hostile.class ...>} \\
\\
This tag instructs the browser to retrieve and run the applet named
{\tt hostile.class} from the server that served this page to the
browser.  The applet that returns is in a format called {\em Java
bytecode}, suitable for running in any JVM.  This bytecode is
subjected to a bytecode verification process, loaded into the
browser's JVM, and executed (see, e.g.,~\cite{LY97}).

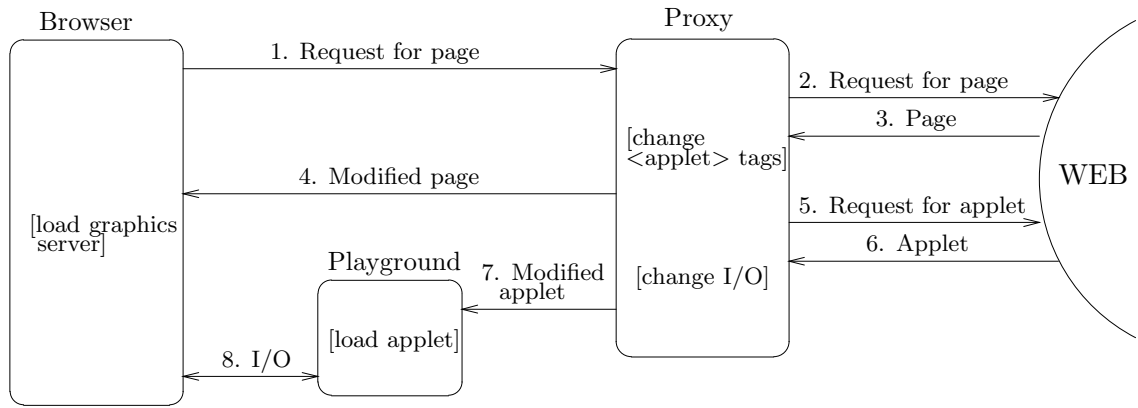
\begin{figure*}[t]
\begin{center}
\input{arch}
\end{center}
\caption{Playground architecture}
\label{approach}
\end{figure*}

In our system, when a browser requests a web page, the request is sent
to a {\em proxy} (Figure~\ref{approach}, step 1).  The proxy forwards
the request to the end server (step 2) and receives the requested page
(step 3).  As the page is received, the proxy parses it to identify
all {\tt <applet>} tags on the returning page, and for each {\tt
<applet>} tag so identified, the proxy replaces the named applet with
the name of a trusted {\em graphics server} applet stored locally to
the browser (i.e., stored in a directory named by the {\tt CLASSPATH}
environment variable).  The proxy then sends this modified page back
to the browser (step 4), which loads the graphics server applet upon
receiving the page.  For each {\tt <applet>} tag the proxy identified,
the proxy retrieves the named applet (steps 5--6) and modifies its
bytecode to use the graphics server in the requesting browser for all
input and output.  The proxy forwards the modified applet to the
playground (step 7), where it is executed using the graphics server in
the browser as an I/O terminal (step 8).

To summarize, there are three important components in our
architecture: the graphics server applet that is loaded into the
user's browser, the proxy, and the playground.  None of these need be
executed on the same machine, and indeed there are benefits to
executing them on different machines (this is discussed in
Section~\ref{security}).  In particular, since untrusted code is
imported into both the proxy and the playground, they should both be
isolated, to the degree possible, from any sensitive resources in the
protected domain (in the limit, they should both be placed outside a
firewall).  The graphics server and the playground are implemented in
Java, and thus can run on any Java compliant environment; the proxy is
a Perl script. The same proxy can be used for multiple browsers and
multiple playgrounds.  In the case of multiple playgrounds, the proxy
can distribute load among playgrounds for improved performance.  In
the following subsections, we describe the functions of these
components in more detail.  Security issues are discussed in
Section~\ref{security}.

\subsection{The graphics server}
\label{graphics-server}

In this section we give an overview of the graphics server that is
loaded into a user's browser in place of an applet provided by a web
server.  Because the graphics server is a Java applet, we must
introduce some Java terminology to describe it.  In Java, a {\em
class} is a collection of data fields and functions (called {\em
methods}) that operate on those fields.  An {\em object} is an
instance of a class; at any point in time it has a state---i.e.,
values assigned to its data fields---that can be manipulated by
invoking the methods of that object (defined by the object's class).
Classes are arranged in a hierarchy, so that a {\em subclass} can
inherit fields and methods from its {\em superclass}.  A running Java
applet consists of a collection of objects whose methods are invoked
by a runtime system, and that in turn invoke one another's methods.
For more information on Java see, e.g.,~\cite{Fla97}.

\subsubsection{Remote AWT classes}
\label{rAWT}

The Abstract Window Toolkit (AWT) is the standard API for implementing
graphical user interfaces (GUI) in Java programs.  The AWT contains classes
for user input and output devices, including buttons, choice boxes,
text fields, images, and a variety of types of windows, to name a few.
Virtually every Java applet interacts with the user by instantiating
AWT classes and invoking the methods of the objects so created.

The intuitive goal of the graphics server is to provide versions of
the AWT classes whose instances can be created and manipulated from
the playground.  For example, the graphics server should enable a
program running on the playground to create a dialog window in the
user's browser, display it to the user, and be informed when the user
clicks the ``ok'' button.  In the parlance of distributed object
technology, such an object---i.e., one that can be invoked from
outside the virtual machine in which it resides---is called a {\em
remote object}, and the class that defines it is called a {\em remote
class}.  So, the graphics server, running in the user's browser,
should allow other machines (the playground) to create and use
``remote AWT objects'' in the user's browser for interacting with the
user.

Accordingly, the graphics server is implemented as a collection of
remote classes, where each remote class (with one exception that is
described in Section~\ref{rApplet}) is a remote version of a
corresponding AWT class.  The (modified) Java applet running on the
playground creates a collection of graphical objects in the graphics
server to implement its GUI remotely, and uses {\em stubs\/} to
interact with them (see Section~\ref{playground}). To minimize the
amount of code in the graphics server, each remote class is a subclass
of its corresponding AWT class, which enables it to inherit many
method implementations from the original AWT class.  Other methods
must be overridden, for example those involving event monitoring: in
the remote class, methods involving the remote object's events (e.g.,
mouse clicks on a remote button object) must be adapted to pass back
the event to the stubs residing in the playground JVM.  In our present
implementation, the remote classes employ Remote Method Invocation
(RMI) to communicate with the playground. RMI is available on all Java
1.1 platforms, including Netscape Communicator, Internet Explorer 4.0,
JDK 1.1, and HotJava 1.0.

\subsubsection{The remote Applet class}
\label{rApplet}

As described in Section~\ref{rAWT}, most classes that comprise the
graphics server are remote versions of AWT classes.  The main
exception to this is a remote version of the {\tt java.applet.Applet}
(or just {\tt Applet}) class, which is the class that all Java applets
must subclass.  The main purpose of the {\tt Applet} class is to
provide a standard interface between applets and their environment.
Thus, the remote version of this class serves to provide this
interface between the applet running on the playground and the
environment with which it must interact, namely the user's browser.

More specifically, this class implements two types of methods.  First,
it provides remote interfaces to the methods of the {\tt Applet}
class, so that applets on the playground can invoke them to interact
with the user's environment.  Second, this class defines a new
``constructor'' method for each remote AWT class (see
Section~\ref{rAWT}).  For example, there is a {\tt constructButton}
method for constructing a remote button in the user's browser.  This
constructor returns a reference to the newly-created button, so that
the remote methods of the button can be invoked directly from the
playground.  Similarly, there is a constructor method for each of the
other remote AWT classes.

When initially started, the graphics server consists of only one
object, whose class is the remote Applet class, called {\tt
BrowserServer.class}.  The applet on the playground can then invoke
the methods of this object (and objects so created) to create the
graphical user interface that it desires for the user.

\subsection{The proxy}
\label{proxy}

The proxy serves as the browser's and playground's interface to the
web.  It retrieves HTML pages for the browser and Java bytecodes for
the playground, and transforms them to formats suitable for the
browser or playground to use.  Like the playground, the proxy can be
placed outside the protected domain (e.g., outside the firewall) to
limit trust in it.

When streaming in an HTML page for the browser, the proxy parses the
returning HTML, identifies all {\tt <applet>} tags in the page, and
replaces them with references to the remote Applet class of the
graphics server (see Section~\ref{rApplet}).  Thus, when the browser
receives the returned HTML page it loads this remote Applet class
(stored locally), instead of the applet originally referenced in the
page.

When retrieving a Java bytecode file for the playground, the proxy
transforms it into bytecode that interacts with the user on the
browser machine while running at the playground.  It does so by
replacing all invocations of AWT methods with invocations of the
corresponding remote AWT methods at the browser, or more precisely,
with invocations of playground-side stubs for those remote AWT methods
(which in turn call remote AWT methods).  This involves parsing the
incoming bytecode and making automatic textual substitutions to change
the names of AWT classes to the names of the representative stubs of
the corresponding remote AWT classes. Though the cost of this function
itself is unnoticeable, diverting all incoming applets through the
proxy may form a bottleneck at the proxy, and hence, this function
could instead be performed at the playground.

\subsection{The playground}
\label{playground}

The playground is a machine that loads modified applets from the proxy
and executes them.  As described above, the proxy modifies the
applet's bytecodes so that playground-resident stubs for remote AWT
methods are called instead of the (non-remote) AWT methods themselves.
So, when a modified applet runs on the playground, a ``skeleton'' of
its GUI containing stubs for corresponding remote graphics objects is
built on the playground. The stubs contain code for remotely invoking
the remote objects' methods at the user's browser and for handling
events passed back from the browser. For example, in the case of a
dialog window with an ``ok'' button, stubs for the window and for
button objects are instantiated at the playground. Calls to methods
having to do with displaying the window and button are passed to the
remote objects at the user's browser, and ``button press'' events are
passed back to methods provided by the button's stub to handle such
events.  These stubs are stored locally on the playground, but aside
from this, the playground is configured as a standard JVM.

A playground is a centralized resource that can be carefully
administered.  Moreover, investments in the playground (e.g.,
upgrading hardware or performing enhanced monitoring) can improve
applet performance and security for all users in the protected domain.
There can even be multiple playgrounds for load-balancing.

\section{Security}
\label{security}

The security goal of our system is to protect resources in the
protected domain from hostile applets that are downloaded by users in
that domain. As mentioned above, this is done by isolating untrusted
applets from the protected resources.\footnote{We limit our attention
to protecting data that users do not offer to hostile applets.
Protecting data that users offer to hostile applets by, e.g., typing
it into the applet's interface, must be achieved via other protections
that are not our concern here (though we can utilize them on our
playground if available).}  In this section we detail how this is
achieved, in three components.  First, we show how our our system
prevents applets from being loaded into user's browsers.  Second, we
describe how private resources in the protected domain are guarded
from access by (even privileged) processes on the playground.  Third,
we show how our system prevents other known attacks that can be
mounted through legitimate use of resources from the playground (e.g.,
denial of service).  We address each of these issues in a separate
subsection below, and then conclude this section with a discussion of
RMI security.

\subsection{Preventing hostile applets from entering the protected domain}
\label{prevent-class-loading}

Achieving strong protection for the domain's private resources relies
on preventing the JVM in the user's browser from loading any classes
from the network (i.e., from outside the {\tt CLASSPATH}).  This can
be achieved in one of two ways in our system.

\paragraph{Trusted proxy} One approach is to depend on the proxy to
intercept and deny entry to any classes destined for protected
machines.  To achieve this, it does not suffice for the proxy to
simply rewrite {\tt <applet>} tags in incoming HTML pages, as it
already does for functional reasons as described in
Section~\ref{arch}.  For example, if the playground passes an object
of an unknown class to the graphics server in the browser (e.g., as a
parameter to a remote method call), or if an {\tt <applet>} tag is not
rewritten by the proxy because it is disguised (e.g., emitted by
JavaScript code in a page, or otherwise encoded), then the browser may
request a class from the network.  In this case, the proxy must
intercept the request or the incoming class, and prevent the class
from reaching the browser. In~\cite{MRR97}, several mechanisms for
filtering class files are discussed, as well as some difficulties. As
mentioned in Section~\ref{related} above, several companies (Finjan,
Security7, eSafe) offer commercial products that identify class files
at the firewall and filter them based on various security policies.
The advantage of this approach is that it works with any browser
``off-the-shelf'': it requires no changes to the browser beyond
specifying the proxy as the browser's HTTP and SSL proxy, which can
typically be done using a simple preferences menu in the browser.  A
disadvantage, however, is that the proxy becomes part of the trusted
computing base of the system, and as shown in~\cite{MRR97},
effectively blocking classes can be costly.  For this reason, the
proxy must be written and maintained carefully, and we refer to this
approach as the ``trusted proxy'' approach.

\paragraph{Untrusted proxy} A second approach to preventing the
browser from loading classes over the network is to directly disable
network class loading in the browser.  The main disadvantage of this
approach is that it requires either configuration or source-code
changes to all browsers in the protected domain.  In particular, for
most popular browsers today (including Netscape Communicator and
Internet Explorer 4.0), a source-code change seems to be required to
achieve this, but we expect such changes to become easier as browser's
security policies become more configurable.  An advantage of this
approach is that it excludes the proxy from the trusted computing base
of the system, and hence we call this the ``untrusted proxy''
approach.  In this approach only the browser and the graphics server
classes are in the trusted computing base. \\

To more precisely show how the above approaches prevent network class
loading by the browser, below we describe what causes classes to be
loaded from the network and how our approaches prevent this.

\begin{level1}

\item Section~\ref{arch} briefly described the process by which a
browser loads an applet specified in an {\tt <applet>} tag in an HTML
page.  As described in Sections~\ref{arch} and~\ref{addressing}, the
proxy rewrites the {\tt code} attribute of each {\tt <applet>} tag to
reference the trusted graphics server applet that is stored locally to
the browser.  If this rewriting fails for some reason (e.g., because
the {\tt <applet>} tag is dynamically emitted by JavaScript code),
then in the untrusted proxy approach, the browser still will not issue
a request for the untrusted applet.  In the trusted proxy approach,
the browser will issue the request, but the proxy will deny the
applet's passage.

\item Once an applet is loaded and started, the {\em applet class
loader} in the browser loads any classes referenced by the applet.  If
a class is not in the core Java library or stored locally (i.e., in
the directory specified by {\tt CLASSPATH}), the applet class loader
would normally retrieve the class over the network.  However, in our
approach, since only the graphics server applet executes in the
browser, and since this applet refers only to other local classes, the
applet class loader will ever need to load only local classes.

\item As described in Section~\ref{arch}, the modified applet on the
playground invokes remote methods in the graphics server.  Because our
present implementation uses RMI to carry out these invocations, the
{\em RMI class loader} can load additional classes to pass parameters
and return values.  As above, the RMI class loader first looks in
local directories to find these classes before going to the network to
retrieve them.  It is possible that the playground applet passes an
object whose class is not stored locally on the browser machine
(particularly if the playground is corrupted).  In the trusted proxy
approach, the RMI class loader goes to the network, via the proxy, to
retrieve the class, but the class is detected and denied by the proxy.
In the untrusted proxy approach, the RMI class loader returns an
exception immediately upon determining that the class is not available
locally.

\end{level1}

\subsection{Isolating untrusted applets}
\label{isolation}

Once untrusted applets are diverted to the playground, security relies
on preventing those applets from accessing protected resources.
Applets on the playground are obviously subject to the security
mechanisms enforced by the JVM. Unfortunately, several bugs in the
type safety mechanisms of Java have provided ways for applets to
bypass Java sandboxes, including some in popular
browsers~\cite{DFW96,MF97}.  These penetrations typically enable the
applet to perform any operation that the operating system allows.  We
anticipate that in the foreseeable future, type safety errors will
continue to exist, and therefore we must presume that applets running
on our playground may run unconstrained by the sandbox.  However, they
are still confined by the playground operating system's protections
and, ultimately, to attack only those resources available to the
playground.  In this section, we described several approaches to limit
what resources are available to processes---and thus to
applets---running on the playground.  We expect that through proper
network and operating system configuration, hostile applets can be
effectively isolated from protected resources.

A first step is for the playground's JVMs (and thus the applets) to
execute under accounts different from actual users' accounts and that
have few permissions associated with them.  For some resources, e.g.,
user files available to the playground, this achieves security
equivalent to that provided by the access control mechanisms of the
playground's operating system.  (Similarly, JVMs on the playground
execute under different accounts, to reduce the threat of inter-JVM
attacks; see Section~\ref{known}.)

Further configuration of the network may provide additional defenses.
For example, if network file servers are configured to refuse requests
from the playground (and if machines' requests are authenticated, as
with AFS~\cite{How88}), then even the total corruption of the
playground does not immediately lead to the compromise of user files.
Similarly, if all machines in the protected domain are configured to
refuse network connections from the playground, except to designated
ports reserved for browsers' graphics servers to listen, then the
compromise of the playground should gain little for the attacker.

In the limit, an organization's playground can be placed outside its
firewall, thereby giving applets no greater access than if they were
run on the servers that served them.  However, because most firewalls
disallow connections from outside the firewall to inside, additional
steps may be necessary so that communication can proceed uninhibited
between the graphics server in the browser and the applet on the
playground.  In particular, RMI in Java 1.1 (used in our prototype)
opens network connections between the browser and the playground in
both directions, i.e., from the browser to the playground and vice
versa.  One approach to enable these connections across a firewall is
to multiplex them over a single connection from the graphics server to
the playground (i.e., from inside to outside).  This can be achieved
if both the graphics server and the playground applet interpose a
customized connection implementation (e.g., by changing the {\tt
SocketImplFactory}), but for technical reasons this does not appear to
be possible with all off-the-shelf browsers (e.g., it appears to work
with HotJava 1.0 but not Netscape 3.0).  Another alternative is to
establish reserved ports on which graphics servers listen for
connections from playground applets, and then configure the firewall
to admit connections from the playground to those ports.

\subsection{Other attacks}
\label{known}

In this section, we review the effect of several remaining, known
types of attacks that applets can mount within their legitimate use of
resources, and describe the extent to which our system can defend
against them.

\paragraph{Denial of service}

In a denial of service attack, a hostile applet might disable or
significantly degrade access to system resources such as the CPU,
disk, network and interactive devices.  Ladue~\cite{Lad96} presents
several such applets, e.g., that consume CPU even after the user
clicks away from the applet origin page, that monopolize system locks,
or that pop up windows on the user's screen endlessly. Using our Java
playground, most of these applets have no effect on the protected
domain, and only affect the playground machine. However, an applet
that pops up windows endlessly causes the graphics server running in
the user's browser to create an infinite stream of windows.
Uncontrolled, this may prevent access to the user terminal altogether
and require that the user reboot her machine or otherwise shut off her
browser.  One approach to defend against this is to configure the
graphics server and/or the playground to limit the number or rate of
window creations. This can be done using several commercially
available products, e.g., due to Finjan and eSafe, and is in fact an
integral part of Digitivity's Cage environment~\cite{Her97}. However,
it is not the focus of our work here.

In another type of denial of service, an applet may deny service to
other applets within the JVM, e.g., by killing off others'
threads~\cite{Lad96}.  Although the sandbox mechanisms of most
browsers are intended to separate applets in different web pages from
one another, several ways of circumventing this separation have been
shown~\cite{DFW96,Lad96}.  This can be prevented in our system if the
applets for each page run in a separate JVM on the playground under a
separate user account, and hence are unable to directly affect applets
from another page (except by attacking the playground itself).

\paragraph{Violating privacy}

The Java security policy in browsers is geared towards maintaining
user privacy by disallowing loaded applets access to any local
information.  In some cases, however, a Java applet can reveal a lot
about a user whose browser executes it.  For example, in~\cite{Lad96},
Ladue presents an applet that uses a {\em sendmail} trick to send mail
on the user's behalf to a sendmail daemon running on the applet's
server.  When this applet is downloaded onto a Unix host (running the
standard {\em ident} service), this mail identifies not only the
user's IP address, but also the user's account name.  In our system,
the applet runs on the playground machine under an account other than
the user's, and the information that it can reveal is limited only to
what is available on the playground.

\subsection{RMI security}
\label{RMI-security}

Even if the protected domain is secured against access from the
playground machine, there is still one way in which hostile applets
might attempt to penetrate protected resources, namely through the RMI
connection established between the playground and the user's browser.
In particular, RMI is a relatively new technology that could
conceivably present new vulnerabilities.  A first step toward securing
RMI is to support authenticated and encrypted transport, so that a
network attacker cannot alter or eavesdrop on communication between
the browser and the playground.  This can also be achieved by
interposing encryption at the object serialization layer
(see~\cite{OSS96}).

A more troubling threat is possible vulnerabilities in the object
serialization routines that are used to marshal parameters to and
return values from remote method invocations.  In the worst case, a
corrupted playground could conceivably send a stream of bytes that,
when unserialized at the browser, corrupt the type system of the JVM
in the browser.  Here our decision to generally pass only primitive
data types (e.g., integers, strings) as parameters to remote method
invocations (see Section~\ref{implementation}) would seem to be
fortuitous, as it greatly limits the number structurally interesting
classes that the attacker has at its disposal for attempting such an
attack.  However, the possibility of a vulnerability here cannot yet
be ruled out, and several research efforts are presently examining RMI
in an effort to identify and correct such problems.  This process of
public scrutiny is one of the main advantages to building our system
from public and widely used components.

\section{Implementation}
\label{implementation}

\subsection{An example}
\label{example}

In order to understand how the components described in
Section~\ref{arch} work together, in this section we illustrate the
execution of a simple applet.  This example describes how the applet
is automatically transformed to interact with the browser remotely,
and how the graphics server and its playground-side stubs interact
during applet execution.  This section necessarily involves low-level
detail, but the casual reader can skip ahead to the next section
without much loss of continuity.

The applet we use for illustration is shown in Figure~\ref{Click}.
This is a very simple (but complete) applet that prints the word
``Click!'' wherever the user clicks a mouse button.  For the purposes
of this discussion, it implements two methods that we care about.  The
first is an {\tt init()} method that is invoked once and registers
{\tt this} (i.e., the applet object) as one that should receive mouse
click events.  This registration is achieved via a call to its own
{\tt addMouseListener()} method, which it inherits from its {\tt
Applet} superclass.  The second method it implements is a {\tt
mouseClicked()} method that is invoked whenever a mouse click occurs.
This method calls {\tt getGraphics()}, again inherited from {\tt
Applet}, to obtain a {\tt Graphics} object whose methods can be called
to display graphics.  In this case, the {\tt mouseClicked()} method
invokes the {\tt drawString()} method of the {\tt Graphics} object to
draw the string ``Click!'' where the mouse was clicked.

\begin{figure}[tbh]
{\footnotesize
\begin{verbatim}
import java.applet.*;
import java.awt.*;
import java.awt.event.*;

public class Click extends Applet
  implements MouseListener {

  public void init() {
    // Tell this applet what MouseListener objects
    // to notify when mouse events occur.  Since we
    // implement the MouseListener interface ourselves,
    // our own methods are called.
    this.addMouseListener(this);
  }

  // A method from the MouseListener interface.
  // Invoked when the user clicks a mouse button.
  public void mouseClicked(MouseEvent e) {
    Graphics g = this.getGraphics();
    g.drawString("Click!", e.getX(), e.getY());
  }

  // The other, unused methods of the MouseListener
  // interface.
  public void mousePressed(MouseEvent e) {;}
  public void mouseReleased(MouseEvent e) {;}
  public void mouseEntered(MouseEvent e) {;}
  public void mouseExited(MouseEvent e) {;}
}
\end{verbatim}
}
\caption{An applet that draws ``Click!'' wherever the user clicks}
\label{Click}
\end{figure}

The {\tt Click} applet is a standalone applet that is not intended to
be executed using a remote graphics display for its input and output.
Thus, when our proxy retrieves (the bytecode for) such an applet, the
applet must be altered before being run on the playground.  For one
thing, the {\tt addMouseListener()} method invocation must somehow be
passed to the browser to indicate that this playground applet wants to
receive mouse events, and the mouse click events must be passed back
to the playground so that {\tt mouseClicked()} is invoked.

In our present implementation, passing this information is achieved
using Java Remote Method Invocation (RMI).  Associated with each
remote class is a stub for calling it that executes in the calling
JVM.  The stub is invoked exactly as any other object is, and once
invoked, it marshals its parameters and passes them across the network
to the remote object that services the request.  Below, the stubs are
described by {\em interfaces} with the suffix {\tt Xface}.  For
example, {\tt BrowserXface} is the interface that is used to call the
{\tt BrowserServer} object of the graphics server.

RMI provides the mechanism to invoke methods remotely, but how do we
get the {\tt Click} applet to use RMI?  To achieve this, we exploit
the subclass inheritance features of Java to interpose our own
versions of the methods it invokes.  More precisely, we alter the {\tt
Click} applet to subclass our own {\tt PGApplet}, rather than the
standard {\tt Applet} class.  This is a straightforward modification
of the bytecode for the {\tt Click} applet.  By changing what {\tt
Click} subclasses in this way, the {\tt addMouseListener()} method
called in Figure~\ref{Click} is now the one in Figure~\ref{PGApplet}.
{\tt PGApplet}, shown in Figure~\ref{PGApplet},\footnote{For
readability, in Figures~\ref{PGApplet}--\ref{BrowserGraphics} we omit
{\tt import} statements, error checking, exception handling ({\tt
try/catch} statements), etc.} passes calls, when necessary, to the
remote applet object of the graphics server described in
Section~\ref{rApplet}.  The {\tt addMouseListener()} method simply
adds its argument (the {\tt Click} applet) to an array of mouse
listeners and, if this is the first to register, registers itself as a
mouse listener at the graphics server.

\begin{figure}[p]
{\footnotesize
\begin{verbatim}
public class PGApplet
  extends Applet implements PGAppletXface {
  BrowserXface bx;
  MouseListener ml_array[] = new MouseListener[...];
  int ml_index = 0;

  // Adds a MouseListener.  If this is the first, then
  // register this object at the graphics server as a
  // MouseListener.
  public synchronized void
         addMouseListener(MouseListener l) {
    ml_array[ml_index++] = l;
    if (ml_index == 1) {
      bx.addPGMouseListener(this);
    }
  }

  // Part of the PGAppletXface remote interface.
  // Invoked from the browser graphics server when the
  // mouse is clicked.
  public void PGMouseClicked(BrowserEventXface e) {
    int i;
    PGMouseEvent pme = new PGMouseEvent(e);
    for (i = 0; i < ml_index; i++)
      ml_array[i].mouseClicked(pme);
  }

  // Returns an object that encapsulates the remote
  // graphics object of the browser applet.
  public Graphics getGraphics () {
    return new PGGraphics(bx.getBrowserGraphics());
  }
  ...
}
\end{verbatim}
}
\caption{Part of the PGApplet class (executes on the playground)}
\label{PGApplet}
\end{figure}

\begin{figure}[p]
{\footnotesize
\begin{verbatim}
public class PGGraphics extends Graphics {

  // Reference to the BrowserGraphics remote object
  // in the graphics server.
  private BrowserGraphicsXface bg;

  // The constructor for this object.  Calls its
  // superclass constructor and saves the reference
  // to the BrowserGraphics remote object.
  public PGGraphics(BrowserGraphicsXface b) {
    super();
    bg = b;
  }

  // Invokes drawString in the graphics server.
  public void drawString(String str, int x, int y) {
    bg.drawString(str, x, y);
  }
  ...
}
\end{verbatim}
}
\caption{Part of the PGGraphics class (executes on the playground)}
\label{PlaygroundGraphics}
\end{figure}

This registration at the graphics server is handled by the {\tt
addPGMouseListener()} remote method of {\tt BrowserServer}, the class
of the remote applet object running on the browser (see
Section~\ref{rApplet}).  The relevant code of {\tt BrowserServer} is
shown in Figure~\ref{BrowserServer}.  Recall that {\tt BrowserServer}
implements the {\tt BrowserXface} {\em interface} that specifies the
remote methods that can be called from the playground.  The {\tt
addPGMouseListener()} method, which is one of those remote methods,
records the fact that the playground applet wants to be informed of
mouse events and then registers its own object as a mouse listener, so
that its object's {\tt mouseClicked()} method is invoked when the
mouse is clicked.  Such an invocation passes the mouse-click
event---or more precisely, a reference to a {\tt BrowserEvent} remote
object that holds a reference to the actual mouse-click event
object---back to the {\tt PGMouseClicked()} remote method of the {\tt
PGApplet} class.  The {\tt PGMouseClicked()} method invokes {\tt
mouseClicked()} with a {\tt PGMouseEvent} object, which holds the
reference to the {\tt BrowserEvent} object.  That is, the {\tt
PGMouseEvent} object translates invocations of its own methods (e.g.,
{\tt getX()} and {\tt getY()} in Figure~\ref{Click}) into invocations
of the corresponding {\tt BrowserEvent} remote methods, which in turn
translates them into invocations of the actual {\tt Event} object in
the browser.  For brevity, the {\tt BrowserEvent} and {\tt
PGMouseEvent} classes are not shown.

The call to {\tt getGraphics()} in {\tt Click} is also replaced with
the {\tt PGApplet} version.  As shown in Figure~\ref{PGApplet}, the
{\tt getGraphics()} method of {\tt PGApplet} retrieves a reference to
a remote {\tt BrowserGraphics} object, via the {\tt
getBrowserGraphics()} method of Figure~\ref{BrowserServer}.  The {\tt
getGraphics()} method of {\tt PGApplet} then returns this {\tt
BrowserGraphics} reference encapsulated within a {\tt PGGraphics}
object for calling it.  So, when {\tt Click} invokes {\tt
drawString()}, the arguments are passed to the browser and executed
(Figures~\ref{PlaygroundGraphics},\ref{BrowserGraphics}).

\begin{figure}[p]
{\footnotesize
\begin{verbatim}
public class BrowserServer
  extends Applet
  implements BrowserXface, MouseListener {

  // Reference to the MouseListener object on
  // the playground
  PGMouseListenerXface ml;

  // Part of the BrowserXface remote interface.  Invoked
  // from the playground to add a remote MouseListener.
  public void
         addPGMouseListener(PGMouseListenerXface pml) {
    ml = pml;
    addMouseListener(this);
  }

  // Invoked whenever the mouse is clicked.  Passes the
  // event to the MouseListener on the playground.
  public void mouseClicked(MouseEvent event) {
    ml.PGMouseClicked(new BrowserEvent(event));
  }

  // Returns a remote object that encapsulates the
  // graphics context of this applet.
  public BrowserGraphicsXface getBrowserGraphics() {
    Graphics g = getGraphics();
    return new BrowserGraphics(g);
  }
  ...
}
\end{verbatim}
}
\caption{Part of the BrowserServer class (executes in the browser)}
\label{BrowserServer}
\end{figure}

\begin{figure}[p]
{\footnotesize
\begin{verbatim}
public class BrowserGraphics
  extends Graphics implements BrowserGraphicsXface {

  private Graphics g;

  // The constructor for this class.  Calls the
  // superclass constructor, saves the pointer to the
  // "real" Graphics object (passed in), and exports
  // its interface to be callable from the playground.
  public BrowserGraphics(Graphics gx) {
    super();
    g = gx;
    UnicastRemoteObject.exportObject(this);
  }

  // Part of the BrowserGraphicsXface remote interface.
  // Invoked from the playground to draw a string.
  public void drawString(String str, int x, int y) {
    g.drawString(str, x, y);
  }
  ...
}
\end{verbatim}
}
\caption{Part of the BrowserGraphics class (executes in the browser)}
\label{BrowserGraphics}
\end{figure}

\subsection{Passing by reference}
\label{reference}

In the example of the previous section, all parameters that needed to
be passed across the network were {\em serializable}.  {\em Object
serialization} refers to the ability to write the complete state of an
object to an output stream, and then recreate that object at some
later time by reading its serialized state from an input
stream~\cite{OSS96,Fla97}.  Object serialization is central to remote
method invocation---and thus to communication between the graphics
server and the playground stubs---because it allows for method
parameters to be passed to a remote method and the return value to be
passed back.  In the example of Section~\ref{example}, the remote
method invocation {\tt bg.drawString(str, x, y)} in the {\tt
PGGraphics.drawString()} method of Figure~\ref{PlaygroundGraphics}
causes no difficulty because each of {\tt str} (a string) and {\tt x}
and {\tt y} (integers) are serializable.

However, not all classes are serializable.  An example is the {\tt
Image} class, which represents a displayable image in a
platform-dependent way.  So, while the previous invocation of {\tt
bg.drawString(str, x, y)} succeeds, a similar invocation {\tt
bg.drawImage(img, x, y, ...)} fails because {\tt img} (an instance of
{\tt Image}) cannot be serialized and sent to the graphics server.
Even if all objects could be serialized, serializing and transmitting
large or complex objects can result in substantial cost.  For such
reasons, an object that can be passed as a parameter to a remote
method of the graphics server is generally constructed {\em in the
graphics server} originally (with a corresponding stub on the
playground).  Then, a reference to this object in the browser is
passed to graphics server routines in place of the object itself.  In
this way, only the object reference is ever passed over the network.

To illustrate this manner of passing objects by reference, we continue
with the example of an {\tt Image}.  When the downloaded applet calls
for the creation of an {\tt Image} object, e.g., via the {\tt
Applet.getImage()} method, our interposed {\tt PGApplet.getImage()}
passes the arguments (a URL and a string, both serializable) to a
remote image creation method in the graphics server.  This remote
method constructs the image, places it in an array of objects, and
returns the array index it occupies.  Playground objects then pass
this index to remote graphics server methods in place of the image
itself.  For example, the {\tt BrowserGraphics} class in the graphics
server implements versions of the {\tt drawImage()} method that accept
image indices and display the corresponding {\tt Image}.

Conversely, there are circumstances in which objects that need to be
passed as parameters to remote methods {\em cannot} first be created
in the browser.  This can be due to security reasons---e.g., the
object's class is a user-defined class that overrides methods of an
AWT class---or because the class of the parameter object is unknown
(e.g., it is only known to implement some interface).  In these
circumstances, a reference is passed in the parameter's place, and
method invocations intended for the object are passed back to the
playground object for processing.  Continuing with our {\tt Image()}
example, such ``callbacks'' can occur when the downloaded applet
applies certain {\em image filters} to an image before displaying it.
One such filter is an {\tt RGBImageFilter}: A subclass of {\tt
RGBImageFilter} defines a per-pixel transformation to apply to an
image by overriding the {\tt filterRGB()} method.  To avoid loading
untrusted code in the browser, such a filter must be executed on the
playground with callbacks to its {\tt filterRGB()} method.

In some circumstances, the need to pass objects by reference can
considerably hurt performance.  Continuing with the {\tt
RGBImageFilter} example above, filtering an image may require that
{\em every image pixel} be passed from the browser to the playground,
transformed by the {\tt filterRGB()} method, and passed back.  This
can result in considerable delay in rendering the image, though our
experience is that this delay is reasonable for images whose pixel
values are indices into a colormap array (i.e., for images that employ
an {\tt IndexColorModel}).

\subsection{Addressing}
\label{addressing}

The previous sections described how an applet running on the
playground is coerced into using the user's browser as its I/O
terminal.  Before any I/O can be performed at the browser, however,
the applet running on the playground and the graphics server running
in the user's browser must be able to find each other to communicate.
This is complicated by the fact that an HTML page can contain any
number of {\tt <applet>} tags that, when modified by the proxy, result
in multiple instances of the graphics server running in the browser.
To retain the intended function of the page, it is necessary to
correctly match each applet running on the playground with its
corresponding instance of the graphics server in the browser.

The addressing scheme that we use requires that the proxy make
additional changes to the HTML page containing applet references prior
to forwarding it to the browser.  Specifically, if the page contains
an {\tt <applet>} tag of the form \\
\\
{\tt <applet code=hostile.class ...>} \\
\\
then the proxy not only replaces {\tt hostile.class} with {\tt
BrowserServer.class} (as described in Section~\ref{arch}), but also
adds a {\em parameter tag} to the HTML page, like this: \\
\\
{\tt <applet code=BrowserServer.class ...>} \\
{\tt <param name=ContactAddress value={\em address}>} \\
\\
Parameter tags are tags that contain name/value pairs.  This one
assigns an {\em address} value, which the proxy generates to be
unique, to be the value of {\tt ContactAddress}.  The {\tt <param>}
tags that appear between an {\tt <applet>} tag and its terminator
({\tt </applet>}, not shown) are used to specify parameters to the
applet when it is run.  In this case, the {\tt BrowserServer.class}
object (i.e., the remote Applet object of the graphics server; see
Section~\ref{rApplet}) looks for the {\tt ContactAddress} field in its
parameters and obtains the address assigned by the proxy.  Once the
{\tt BrowserServer} object is initialized and prepared to service
requests from the playground, it binds a remote reference to itself to
the address assigned by the proxy; this binding is stored in an RMI
name server~\cite{RMIS97}.

The proxy remembers what address it assigned to each {\tt <applet>}
tag and provides this address to the playground in a similar fashion.
That is, the proxy loads applets into the playground by sending to the
playground an HTML page with identical {\tt ContactAddress <param>}
tags to what it forwarded to the browser (for simplicity, this step is
not shown in Figure~\ref{arch}).  A JVM on the playground loads the
referenced applets (via the proxy) and uses the corresponding {\tt
<param>} tags provided with each to look up the corresponding graphics
servers in the RMI name server.

\section{Transparency}
\label{transparency}

In our experience, our system is transparent to users for most
applets.  There are, however, applets for which our playground
architecture is not transparent, and indeed our system may be unable
to execute certain applets at all.  In particular, the remote
interface supported by the graphics server supports the passage of
certain classes as parameters and return values of its remote methods.
If the code running on the playground attempts to pass an object
parameter whose class is an unknown subclass of the expected parameter
class, then the browser is required to load that subclass to
unserialize that parameter.  However, because class loading from the
network is prevented in our system (see
Section~\ref{prevent-class-loading}), the load does not complete and
an exception is generated.  In addition, our prototype presently does
not offer transparent execution for applets that invoke methods by
reflection~\cite[Ch. 12]{Fla97}.  At the time of this writing,
however, the number of applets that cannot be supported due to these
limitations does not seem significant.

A more subtle limitation in the transparency of our approach is that
by moving the applet away from the machine on which the user's browser
executes, the applet's I/O incurs the overhead of communicating over
the network. Our research prototype is not intended for production use
and hence is not optimized for best performance. Nevertheless, we
observe that no significant delays are visible for applets with
low-intensity graphics, e.g., a clock applet or an interactive drawing
tool.

Experience with commercial tools that realize similar architectures
for remote Java windowing, e.g., IBM's RAWT and Citrix's Cage,
indicates several issues that are crucial for the performance of our
approach. First, the serialization of objects that are passed between
the playground and the graphics server are costly operations and can
often be avoided.  Second, the RMI protocol incurs certain overheads,
e.g., performing all remote calls synchronously, that a native
protocol can avoid.  By avoiding unnecessary serialization and
synchrony, these companies have reported significant speedups with
their proprietary graphics-aware protocols, which are implemented
directly on top of TCP/IP, compared with their initial RMI
implementations.  For example, IBM reports performance comparable to,
and sometimes surpassing, that of X Windows~\cite{RBW99}, and Citrix
representatives have claimed performance superior to X.  Similar
optimizations to our playground architecture will be necessary to
support applets that generate heavy graphics traffic, such as animated
applets.

\section{Other mobile code technologies}
\label{1.2}

Java is not the only mobile code technology that raises security
issues for the machines that execute it.  For example, security risks
are also associated with ActiveX controls and web scripting languages
(e.g.,~\cite{AM98}), and with interactions between such technologies:
for example, it is possible for JavaScript code running in a browser
to invoke Java methods through Netscape's LiveConnect tool.  Even
changes to Java itself raise new challenges.  Since the initial work
for the conference version of this paper~\cite{MRR98}, a new version
of Java (1.2) has been shipped and is now being supported by both
Netscape and Microsoft. One of the main differences in
Java 1.2 is that it provides a significantly richer and more flexible
security model for the JVM.  In this section, we briefly examine these
other mobile code technologies and consider the extent to which our
playground architecture impacts them and, where possible, can be
modified to support them and to enhance their security.

\paragraph{Java 1.2} The security model in Java 1.2 provides the
mechanism for defining protected target resources that can be accessed
only by classes signed by certain trusted principals.  As in the
sandbox model, protection of target resources, such as code for
accessing the file system, is provided by the SecurityManager. Unlike
the sandbox model, the user controls the security policy that guards
the use of protected targets.  A user can also define new targets to
be protected.  For any protected target, the user determines a set of
trusted principals that are allowed to access it. Thus, a security
policy can be specified as an access control matrix: on one dimension
are protected targets and on the other are trusted principals; entries
indicate which principals are allowed to access each target.  In order
to realize its privilege, code must carry the digital signature of a
trusted principal.

By physically isolating applets on the playground, our system coarsely
denies access to protected resources.  The challenge in supporting
Java 1.2 flexibility in our system is thus to enable applets with
appropriate signatures to access the resources allowed by the user
while still guarding all protected resources, especially from unsigned
applets.  A viable solution is to route applets containing certain
credentials to one ``trusted'' playground, and other applets to
another playground, and then support remote access from the trusted
playground to certain resources, according to the security policy.  In
the limit, the ``trusted playground'' could be the user's browser
itself, so that applets containing certain credentials would be
allowed to run in the browser (subject to the browser's enforcement of
the user's security policy).  Note that this solution can also
be used when trust in an applet is ensured by other means, e.g., due
to the host that served it.

To implement this solution, the user's security policy is compiled to
produce a list of trusted principals for the proxy.  Code signed by
any principal in this list in permitted access by the user to certain
sensitive resources, and accordingly, is allowed to run on the trusted
playground or the user's browser, according to a centralized security
policy.  Note that it stands to reason to trust this code to run
inside the domain, and even in the user's browser, since it is signed by a
principal trusted by the user (regardless of the specific target for
which it is authorized).  The proxy continues routing all unsigned
applets to the (untrusted) playground.

An applet may initially contain several class files (in a {\em jar\/}
file), and furthermore, may require to load additional classes from
the network at runtime (see description in
Section~\ref{prevent-class-loading}). Hence, the security policy for a
site should also determine whether every class must be properly signed
to run in the protected domain, or whether it suffices for the top
class of an applet to be signed to allow all referred classes to be
loaded into the domain.

\paragraph{ActiveX}
An ActiveX control is a program that could be written in any language,
and could attempt access to any system resource available to
users. Certain browsers, e.g., Intenet Explorer and Netscape, support
importing ActiveX controls embedded in HTML pages, provided that these
controls are signed by authorized principals, as specified by a local
security policy. Once an ActiveX control is imported by a browser, it
is spawned off as a separate process and runs unrestricted. The
security of ActiveX hinges on the complete trust placed on its source;
if an ActiveX control is malicious, the user's machine is seriously
compromised.

Protecting resources from hostile ActiveX controls by running them
remotely on a playground machine should be effective, from a security
point of view.  However, since the authors of ActiveX controls, unlike
authors of Java 1.1 applets, anticipate the control having
unrestricted access to the user's machine, it is likely that the
percentage of legitimate ActiveX controls that take advantage of that
access is higher.  Consequently, the percentage of controls whose
functionality is severely hindered by running on a resource-sanitized
playground would be higher than for applets.

\paragraph{JavaScript}
JavaScript is a scripting language that can be embedded in any HTML
page and executed in browsers such as Netscape and IE. When loaded by
a browser, a JavaScript script within an HTML page can access a
hierarchy of objects reflecting the layout of the page in the browser,
such as windows and frames, as well as various browser properties,
e.g., version and user history.  JavaScript does not include any file
I/O or networking capabilities, and hence, the security threats
associated with it are typically focused on spoofing attacks, e.g.,
through changing the appearances of various parts of the
browser~\cite{FBDW97}, and attacks on the user's privacy, such as
tracking her browsing history~\cite{AM98}.  These attacks are mounted
through legitimate (albeit unplanned) use of its features.

Recent releases of JavaScript attempted to enforce various security
measures, specifically denying a script access to JavaScript objects
of a page from any domain other than the script's origin domain.
Although we are not aware of any break of these security measures, it
is quite possible that they are not effectively enforced.  To prevent
attacks of one script on another, it is conceivable that different
scripts could be executed in different address spaces, thereby
leveraging operating system protections.  If this were done on a
playground similar to ours, then external mechanisms outside the
JavaScript language would be needed to facilitate interaction between
a script running on the playground and the browser.  For example,
Netscape's LiveConnect tool (see below) could be employed to
communicate JavaScript requests from the playground, through Java, to
the user's browser, and similarly back.

\paragraph{LiveConnect}
LiveConnect is a technology that facilitates communication between
Java and JavaScript (and plug-ins).  Applets can interact with
JavaScript through LiveConnect's special classes, {\tt JSObject} and
{\tt JSException}. Likewise, JavaScript can invoke any public applet
methods and pass them values that are properly wrapped by LiveConnect.
We expect that it is possible to support LiveConnect even when Java
applets are running remotely on the playground (and JavaScript does
not), much like it is possible to support interaction between an
applet on the playground and the browser through the Applet object.
Thus, this would require splitting LiveConnect's special object
classes in the way described in Section~\ref{implementation}.

\section{Conclusion}
\label{conclusion}

This paper presented a novel approach to protecting hosts from mobile
code and an implementation of this approach for Java 1.1 applets.  The
idea behind our approach is to execute mobile code in the sanitized
environment of an isolated machine (a ``playground'') while using the
user's browser as an I/O terminal.  We gave a detailed account of the
technology to allow transparent execution of Java applets separately
from their graphical interface at the user's browser.  Using our
system, users can enjoy applets downloaded from the network, while
exposing only the isolated environment of the playground machine to
untrusted code.  Although we presented the playground approach and
technology in the context of Java 1.1, other mobile-code platforms may
also utilize it.

\subsection*{Acknowledgements}

We thank Avi Rubin for many useful discussions.
We are grateful to Drew Dean, Ed Felten, Li Gong and the anonymous
referees of the 1998 IEEE Symposium on Security and Privacy and {\em
IEEE Transactions on Software Engineering} for helpful comments.

\end{document}

%% file: arch.tex
\setlength{\unitlength}{0.00066667in}
\begingroup\makeatletter\ifx\SetFigFont\undefined%
\gdef\SetFigFont#1#2#3#4#5{%
  \reset@font\fontsize{#1}{#2pt}%
  \fontfamily{#3}\fontseries{#4}\fontshape{#5}%
  \selectfont}%
\fi\endgroup%
{\renewcommand{\dashlinestretch}{30}
\begin{picture}(8795,3125)(0,-10)
\put(3687,987){\makebox(0,0)[lb]{\smash{{{\SetFigFont{9}{10.8}{\rmdefault}{\mddefault}{\updefault}7. Modified}}}}}
\put(3762,840){\makebox(0,0)[lb]{\smash{{{\SetFigFont{9}{10.8}{\rmdefault}{\mddefault}{\updefault}    applet}}}}}
\put(162,1362){\makebox(0,0)[lb]{\smash{{{\SetFigFont{9}{10.8}{\rmdefault}{\mddefault}{\updefault}[load graphics}}}}}
\put(162,1210){\makebox(0,0)[lb]{\smash{{{\SetFigFont{9}{10.8}{\rmdefault}{\mddefault}{\updefault}  server]}}}}}
\put(4812,2037){\makebox(0,0)[lb]{\smash{{{\SetFigFont{9}{10.8}{\rmdefault}{\mddefault}{\updefault}[change}}}}}
\put(4812,1887){\makebox(0,0)[lb]{\smash{{{\SetFigFont{9}{10.8}{\rmdefault}{\mddefault}{\updefault}$<$applet$>$ tags]}}}}}
\put(9432.000,1774.500){\arc{2791.008}{2.0513}{4.2319}}
\path(1362,2637)(4737,2637)
\path(4617.000,2607.000)(4737.000,2637.000)(4617.000,2667.000)
\path(8037,2112)(6087,2112)
\path(6207.000,2142.000)(6087.000,2112.000)(6207.000,2082.000)
\path(4737,1662)(1362,1662)
\path(1482.000,1692.000)(1362.000,1662.000)(1482.000,1632.000)
\put(4857,507){\arc{240}{1.5708}{3.1416}}
\put(4857,2749){\arc{240}{3.1416}{4.7124}}
\put(5967,2749){\arc{240}{4.7124}{6.2832}}
\put(5967,507){\arc{240}{0}{1.5708}}
\path(4737,507)(4737,2749)
\path(4857,2869)(5967,2869)
\path(6087,2749)(6087,507)
\path(5967,387)(4857,387)
\path(4737,762)(3537,762)
\path(3657.000,792.000)(3537.000,762.000)(3657.000,732.000)
\path(1362,237)(2412,237)
\path(1482.000,267.000)(1362.000,237.000)(1482.000,207.000)
\path(2292.000,207.000)(2412.000,237.000)(2292.000,267.000)
\put(117,117){\arc{210}{1.5708}{3.1416}}
\put(117,2757){\arc{210}{3.1416}{4.7124}}
\put(1257,2757){\arc{210}{4.7124}{6.2832}}
\put(1257,117){\arc{210}{0}{1.5708}}
\path(12,117)(12,2757)
\path(117,2862)(1257,2862)
\path(1362,2757)(1362,117)
\path(1257,12)(117,12)
\path(6087,2412)(7587,2412)(8187,2412)
\path(8067.000,2382.000)(8187.000,2412.000)(8067.000,2442.000)
\path(8187,1137)(6087,1137)
\path(6207.000,1167.000)(6087.000,1137.000)(6207.000,1107.000)
\put(2517,192){\arc{210}{1.5708}{3.1416}}
\put(2517,882){\arc{210}{3.1416}{4.7124}}
\put(3432,882){\arc{210}{4.7124}{6.2832}}
\put(3432,192){\arc{210}{0}{1.5708}}
\path(2412,192)(2412,882)
\path(2517,987)(3432,987)
\path(3537,882)(3537,192)
\path(3432,87)(2517,87)
\path(6087,1437)(8037,1437)
\path(7917.000,1407.000)(8037.000,1437.000)(7917.000,1467.000)
\put(2262,1737){\makebox(0,0)[lb]{\smash{{{\SetFigFont{9}{10.8}{\rmdefault}{\mddefault}{\updefault}4. Modified page}}}}}
\put(2037,2712){\makebox(0,0)[lb]{\smash{{{\SetFigFont{9}{10.8}{\rmdefault}{\mddefault}{\updefault}1. Request for page}}}}}
\put(1662,312){\makebox(0,0)[lb]{\smash{{{\SetFigFont{9}{10.8}{\rmdefault}{\mddefault}{\updefault}8. I/O}}}}}
\put(4887,958){\makebox(0,0)[lb]{\smash{{{\SetFigFont{9}{10.8}{\rmdefault}{\mddefault}{\updefault}[change I/O]}}}}}
\put(237,2937){\makebox(0,0)[lb]{\smash{{{\SetFigFont{10}{12.0}{\rmdefault}{\mddefault}{\updefault}Browser}}}}}
\put(5112,2966){\makebox(0,0)[lb]{\smash{{{\SetFigFont{10}{12.0}{\rmdefault}{\mddefault}{\updefault}Proxy}}}}}
\put(6687,1212){\makebox(0,0)[lb]{\smash{{{\SetFigFont{9}{10.8}{\rmdefault}{\mddefault}{\updefault}6. Applet}}}}}
\put(6762,2187){\makebox(0,0)[lb]{\smash{{{\SetFigFont{9}{10.8}{\rmdefault}{\mddefault}{\updefault}3. Page}}}}}
\put(2487,462){\makebox(0,0)[lb]{\smash{{{\SetFigFont{9}{10.8}{\rmdefault}{\mddefault}{\updefault}[load applet]}}}}}
\put(8187,1737){\makebox(0,0)[lb]{\smash{{{\SetFigFont{11}{13.2}{\rmdefault}{\mddefault}{\updefault}WEB}}}}}
\put(6162,1512){\makebox(0,0)[lb]{\smash{{{\SetFigFont{9}{10.8}{\rmdefault}{\mddefault}{\updefault}5. Request for applet}}}}}
\put(6162,2487){\makebox(0,0)[lb]{\smash{{{\SetFigFont{9}{10.8}{\rmdefault}{\mddefault}{\updefault}2. Request for page}}}}}
\put(2487,1062){\makebox(0,0)[lb]{\smash{{{\SetFigFont{10}{12.0}{\rmdefault}{\mddefault}{\updefault}Playground}}}}}
\end{picture}
}